# Light-Induced Spin Slanting in 2D Multiferroic Magnet

Jiangyu Zhao[1], Yangyang Feng[1], Kaiying Dou[1], Xinru Li[1], Ying Dai*[,1], Baibiao Huang[1], Yandong Ma*[,1]

[1]School of Physics, State Key Laboratory of Crystal Materials, Shandong University, Shandanan Street 27, Jinan 250100, China

*Corresponding author: daiy60@sdu.edu.cn (Y.D.); yandong.ma@sdu.edu.cn (Y.M.)

**Abstract**

Controlling spin orientation of two-dimensional (2D) materials has emerged as a frontier of condensed-matter physics, resulting in the discovery of various phases of matter. However, in most cases, spin orientation can be stablished only at specific directions of out-of-plane and in-plane, which is a drawback compared with three-dimensional systems, limiting exploration of novel physics. Here, we introduce a methodology for manipulating spin slanting in 2D multiferroic materials through ultrafast pulses of light. Based on model analysis, we find that simultaneous triggering spin-orbit coupling induced interactions from in-plane and out-of-plane orbitals can generate spin slanting. By choosing 2D multiferroic materials with specific low-energy composition endowed by symmetry, such triggering can be readily achieved through ultrafast light illumination, leading to light-induced spin slanting. Using real-time time-dependent density-functional theory, we demonstrate this approach in multiferroic single-layer $CuCr_2Se_4$. This study provides an efficient way to manipulate spin orientation in 2D materials and establishes a general platform to explore physics and applications associated with spin slanting.

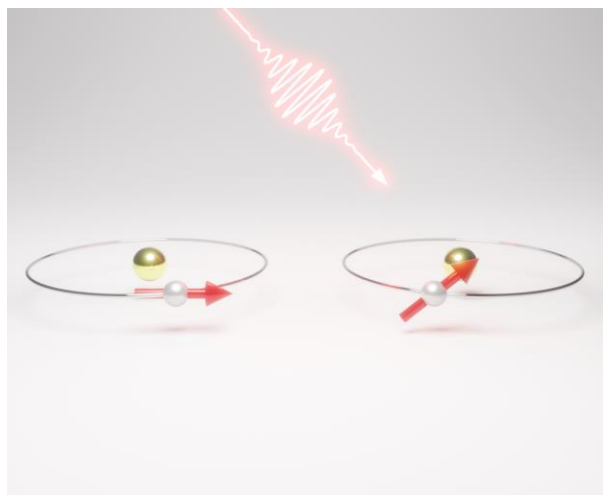

## 1. Introduction

Since the experimental discovery of long-range magnetic order in two-dimensional (2D) $CrI_3$ and $Cr_2Ge_2Te_6$[1,2], 2D magnets have been intensively investigated because of their potential applications in constructing atomically thin spintronic devices[3–15]. Meanwhile, due to the reduced dimensionality, various intriguing phenomena and effects are identified in 2D magnets, which provides an ideal playground to explore fundamental physics. An important physical quantity describing the magnetic nature in 2D magnets is spin orientation, which essentially correlates to the physical properties and novel effects[16–22]. For example, in most 2D topological magnetic systems[23,24], while in-plane spin orientation enforces trivial semiconducting character, out-of-plane spin orientation guarantees quantum anomalous Hall effect; as ferro-valleytricity is forbidden with in-plane spin orientation, it is allowed with out-of-plane spin orientation[25]. Such intimate relationship between exotic physics and spin orientation in 2D materials ignites the ongoing search for efficient control of spin orientation[4,26].

Despite the surge interest in modulating spin orientation, it is generally believed that the preferred directions for spins in most of the 2D lattice are limited to specific directions, i.e., out-of-plane and in-plane[27–30]. Physically, such restricting is attributed to the special coupling between electronics orbitals and spins, which succumbs to the lattice symmetry constrained by reduced dimensionality[31]. This is in sharp contrast to three-dimensional magnetic systems, wherein spins, in principle, can be set along arbitrary directions[32]. Such drawback compared with three-dimensional systems severely restricts the exploration of potential novel physics associated with spin orientation in 2D magnetic systems, posing an outstanding challenge for the development of 2D magnetism. Currently, how to modulate spins in 2D systems along directions beyond in-plane and out-of-plane is an open question. Undoubtedly, once achieved, it would make a significant step toward novel physics and high-performance spintronic devices based on 2D magnetism.

In this letter, we propose a mechanism of modulating spins along directions beyond in-plane and out-of-plane in 2D multiferroic materials based on the framework of light-induced magnetization dynamics. Using straightforward perturbation theory analysis, we show that the competition between spin-orbit coupling (SOC)-induced interactions from in-plane and out-of-plane orbitals that prefer different spin orientations can result in spin slanting. Through choosing 2D multiferroic materials with specific low-energy orbital occupations endowed by symmetry, this competition can be introduced by ultrafast light illumination as optical excitation can alter the orbital occupation, thereby giving rise to light-induced spin slanting. With the help of real-time time-dependent density-functional theory (rt-TDDFT), we further demonstrate this mechanism is attainable in multiferroic single-layer (SL) $CuCr_2Se_4$. Our study offers an unprecedent opportunity to explore potential physics and applications associated with non-traditional spin orientations in 2D magnetic systems.

## 2. Results and Discussion

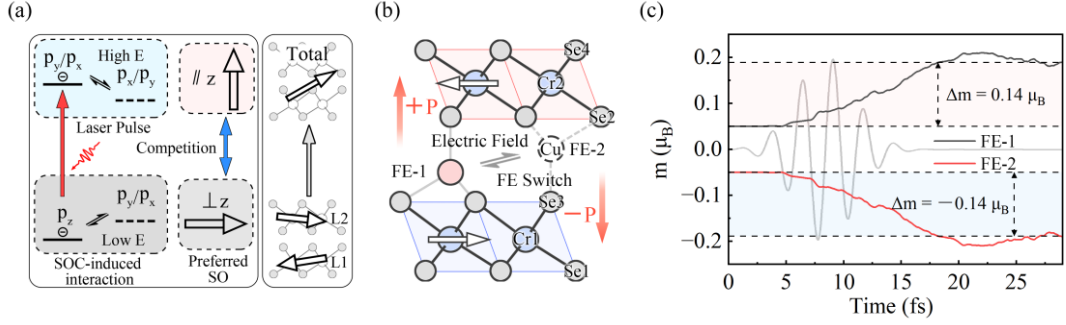

**Figure 1.** (a) Schematic diagram of preferred spin orientations (SO) of magnetic atoms associated with the SOC-induced interactions. In the left panel, the solid and dashed horizontal lines represent the occupied and unoccupied states, respectively. The high/low E marks the high/low energy state. In the right panel, L1 and L2 mark the lower and upper layers, respectively. (b) Crystal structure of SL CuCr$_2$Se$_4$. Red arrows in (b) represent the direction of FE dipole. White arrows in (a) and (b) represent spin orientations. (c) Time-dependent dynamics of total magnetic moments of two ferroelectricity states. Gray line in (c) represents the absolute value of the vector potential of laser pulse.

Physically, spin orientation of 2D materials is closely related to the coupling between electronic orbitals and spin, i.e., the SOC-induced interaction[33,34]. The SOC Hamiltonian can be written as:

$$\hat{H}_{SOC} = \lambda \hat{S} \cdot \hat{L} = \hat{H}_{SOC}^0 + \hat{H}_{SOC}' \tag{1a}$$

$$\hat{H}_{SOC}^0 = \lambda \hat{S}_{z'}(\hat{L}_z \cos\theta + \hat{L}_x \sin\theta \cos\Phi + \hat{L}_y \sin\theta \sin\Phi) \tag{1b}$$

$$\hat{H}_{SOC}' = \frac{\lambda}{2}(\hat{S}_{+'} + \hat{S}_{-'})(-\hat{L}_z \sin\theta + \hat{L}_x \cos\theta \cos\Phi + \hat{L}_y \cos\theta \sin\Phi) \tag{1c}$$

Here, $\lambda$ is SOC parameter, $\hat{L}$ is orbital angular momentum operator, $\hat{S}$ is spin operator, and the subscripts $+/-$ mark raising/lowering operator. $(x, y, z)$ and $(x', y', z')$ are coordinates for $\hat{L}$ and $\hat{S}$, respectively. $\theta$ and $\Phi$ are polar coordinates to describe the spin orientation. Among all the SOC-induced interactions, the dominated ones are the couplings between the occupied state $\psi_i$ and unoccupied state $\psi_j$ with the same spin state. Therefore, we only focus on the SOC-induced interaction $\langle\psi_i|\hat{H}_{SOC}^0|\psi_j\rangle$. Taking $p$ orbital system as an example, the following relationship can be obtained

$$\langle\psi_i|\hat{H}_{SOC}^0|\psi_j\rangle \propto \begin{cases} \cos\theta, \text{interaction between } p_x \text{ and } p_y \\ \sin\theta, \text{interaction between } p_x/p_y \text{ and } p_z \end{cases} \tag{2}$$

Note that SOC-induced interaction lowers the total system energy, spin parallel to out-of-plane direction is favorable for the interaction between in-plane orbitals $p_x$ and $p_y$, while in-plane spin direction is preferred for the interaction between out-of-plane orbital $p_z$ and in-plane orbitals $p_x/p_y$. Similar conclusion can be obtained for d orbitals. For more detail, please see **Table S1**. As a result, controlling spin direction can be actively pursued through manipulation of orbital degrees of freedoms.

In view of the crystal symmetry of 2D lattice, only one of the two interactions shown in **Eq. (2)** can exist in 2D magnets. In this regard, the preferred direction for spins in 2D systems is limited to either out-of-plane or in-plane. To generate spin slanting, one possible way is to introduce the two interactions shown in **Eq. (2)** simultaneously. Once introduced, the competition between SOC-induced interactions from in-plane and out-of-plane orbitals can slant the spin direction. To be more specific, we take one example with the low-energy band shown in the lower region of left panel in **Figure 1(a)** to illustrate this mechanism. In the ground state, the $p_z$ orbital is occupied, while the $p_x$ and $p_y$ orbitals are unoccupied. Owing to the SOC-induced interaction $\langle p_{x/y}|\hat{H}_{SOC}^0|p_z\rangle$, the system prefers in-plane spin orientation. As depicted in the upper region of left panel in **Figure 1(a)**, when some electrons are populated to higher states, the SOC-induced interaction $\langle p_{x/y}|\hat{H}_{SOC}^0|p_{y/x}\rangle$ between higher occupied states ($p_x/p_y$) and unoccupied states ($p_y/p_x$) favors out-of-plane spin orientation. As shown in the right region of left block in **Figure 1(a)**, the additional introduction of $\langle p_{x/y}|\hat{H}_{SOC}^0|p_{y/x}\rangle$ dictate spins to deviate from original orientation, resulting in spin slanting.

Since our target is to introduce the coexistence of $\langle p_{x/y}|\hat{H}_{SOC}^0|p_z\rangle$ and $\langle p_{x/y}|\hat{H}_{SOC}^0|p_{y/x}\rangle$, the carriers should be distributed in both the conduction and valence bands simultaneously. This requirement naturally points to ultrafast light illumination, i.e., using light to induce spin slanting [**Figure 1(a)**]. This is different from the Fermi energy shifting induced by carrier doping, strain or electric field that is usually employed to tune the magnetic anisotropy[19,35], which would result in the creation of $\langle p_{x/y}|\hat{H}_{SOC}^0|p_z\rangle$ and annihilation of $\langle p_{x/y}|\hat{H}_{SOC}^0|p_{y/x}\rangle$, rather than their coexistence. To realize light-induced spin slanting shown in left panel of **Figure 1(a)**, two magnetic sublattice should be separated in space. This can promote the formation of spin transfer, thereby enhancing the system's response to light. For facilizing the introduction of ultrafast light pulses, the inversion symmetry of the system should be broken. Otherwise, the identical responses of two equivalent magnetic atoms to light would cancel each other out. Inspired by these insights, 2D ferroelectric (FE) ferrimagnetic materials are the ideal candidate systems, and the corresponding schematic light-induced spin slanting is depicted in the right panel of **Figure 1(a)**.

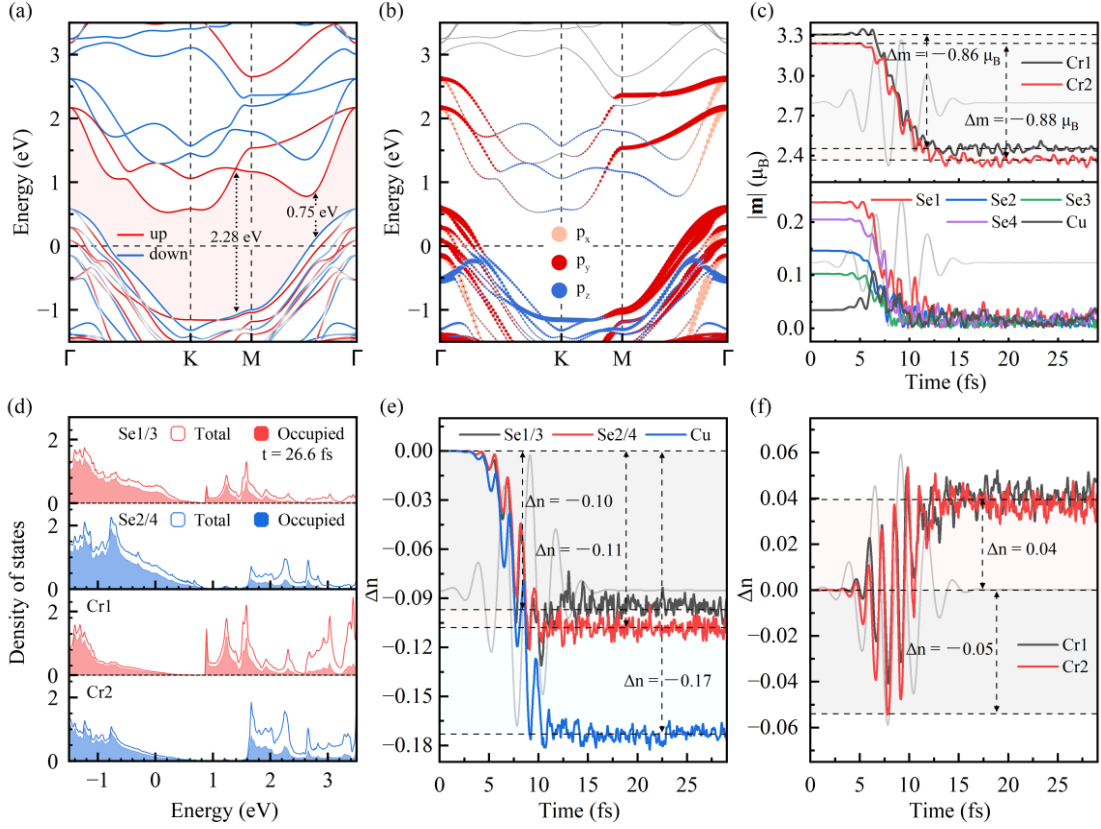

**Figure 2.** (a) Spin-resolved and (b) orbital-resolved band structures of SL $CuCr_2Se_4$ under FE-1 with SOC. (c) Time evolution of local magnetic moment on each atom under laser excitation for SL $CuCr_2Se_4$ under FE-1. (d) Orbital-resolved density of occupied states after laser excitation for SL $CuCr_2Se_4$ under FE-1. The lines indicate the density of both occupied and unoccupied states and the fill color region indicates the density of occupied states. (e) Time evolution of charge transfer of Se and Cu atoms. (f) Time evolution of charge transfer of Cr atoms. In (a), (b) and (d), the Fermi energy is shifted to zero. Gray line in the background represents the absolute value of the vector potential of laser pulse.

One candidate system for realizing this mechanism is SL $CuCr_2Se_4$, which has been synthesized in experiment[36–41] and extensively studied in previous works[42]. It should be emphasized that SL $CuCr_2Se_4$ has been experimentally confirmed to be with ferroelectric multiferroicity[40]. And the metallic ferroelectricity is also theoretically proposed and experimentally confirmed in many recent works[40,43–47]. The crystal structure of SL $CuCr_2Se_4$ is depicted in **Figure 1(b)**. It has a hexagonal lattice and belongs to P3m1 (No. 156) space group. The lattice parameter for SL $CuCr_2Se_4$ is optimized to 3.63 Å. Its unit cell consists of two $CrSe_2$, which are separated by an off-centered Cu atom. The displacement of Cu atom leads to two energetically degenerate FE states (i.e., FE-1 and FE-2). The electric polarization is calculated to be 3.47 pC/m.

SL $CuCr_2Se_4$ is spin polarized. Its magnetic ground state is shown in **Figure 1(b)**. SL $CuCr_2Se_4$ prefers in-plane spin orientation, with a magnetic anisotropy energy (MAE) of -0.4 meV/unit cell. The existence of ferroelectricity breaks the equivalence between Cr1 and Cr2 atoms, leading to different magnetic moments. In detail, in FE-1, the magnetic moments contributed by Cr1 and Cr2 atoms are calculated to be 3.31 $\mu_B$ and -3.24 $\mu_B$, respectively, and the net magnetic moment is 0.05 $\mu_B$ per unit cell. When transforming into FE-2 under an applied electric field, the environments of Cr1 and Cr2 are exchanged, which results in magnetic moments of 3.24 $\mu_B$ and -3.31 $\mu_B$, respectively. This exchange causes a reversal in direction of total magnetic moment, thereby demonstrating the coupling between magnetism and ferroelectricity in SL $CuCr_2Se_4$.

Then, we investigate the magnetization dynamics of SL $CuCr_2Se_4$ under ultrafast laser pulse. For convenience of discussion, if otherwise not specified, we focus on FE-1 in the following. **Figure 2(a)** shows the band structure of SL $CuCr_2Se_4$. The gap width between top valance band and lowest conduction band [the region filled with red color in **Figure 2(a)**] ranges from 0.74 eV to 2.24 eV. In light of this fact, we first employ an ultra-short laser pulse with a wavelength of 800 nm, corresponding to a photon energy of 1.55 eV, which is a commonly used wavelength for Ti:sapphire lasers and close to the averaged band gap. In rt-TDDFT simulations, the laser pulse is modeled with a Gaussian envelope, having intensity of $1\times10^{13}$ w/cm$^2$ and pulse width of 16 fs, which corresponds to an energy density of 23.1 mJ/cm$^2$ and corresponds to a peak electric field strength of $8.68\times10^9$ V/m. The vector potential oscillates along x-direction. **Figure 1(c)** displays the time evolution of the total magnetic moment per unit cell, from which we find faster magnetization process. This results in the change of total magnetic moment of $\Delta m = \sim 0.14$ $\mu_B$ per unit cell. The time scale of magnetization corresponds to the duration of the imposed laser pulse, which is similar to the cases of $MnSe_2$ and $RuCl_3$ [28,48]. Notably, the coupling between ferroelectricity and magnetism allows the FE control of magnetization dynamics of SL $CuCr_2Se_4$. Specifically, as shown in **Figure 1(c)**, the total magnetic moment increases in the opposite direction for FE-2.

To gain deeper understanding of the magnetization dynamics, we study the time evolution of magnetic moment for each atom. As shown in the upper panel of **Figure 2(c)**, both Cr1 and Cr2 atoms exhibit considerable demagnetization in response to laser pulse. And the demagnetization nature between Cr1 and Cr2 atoms is different, which can be attributed to the existence of ferroelectricity that disrupts the equivalence of two Cr atoms. In fact, the demagnetization mainly correlates to the spin transfer between the two Cr atoms. Specifically, in quantitative comparison of the demagnetization on two Cr atoms, the magnetic moment on Cr1 decreases by 0.86 $\mu_B$, while the magnetic moment on Cr2 decreases by 0.88 $\mu_B$. In addition, as shown in the lower panel of **Figure 2(c)**, the small magnetic moments on Se and Cu atoms are reduced to nearly zero. This suggests the spin transfer also occurs from Se1 and Se3 to Cr1 atoms, and from Se2 and Se4 to Cr2 atoms. For more information, the time evolutions of charge transfer on each atom are shown in **Figure 2(e)** and **2(f)**.

**Figure 2(d)** presents the orbital-resolved density of occupied states after laser excitation for SL $CuCr_2Se_4$. It is worth mentioning that, the states between 0 and 0.5 eV are filled by the electrons excited from deeper states. Obviously, the carriers are distributed in both conduction and valence bands simultaneously under ultrafast laser pulse, which satisfies the requirement of the proposed mechanism.

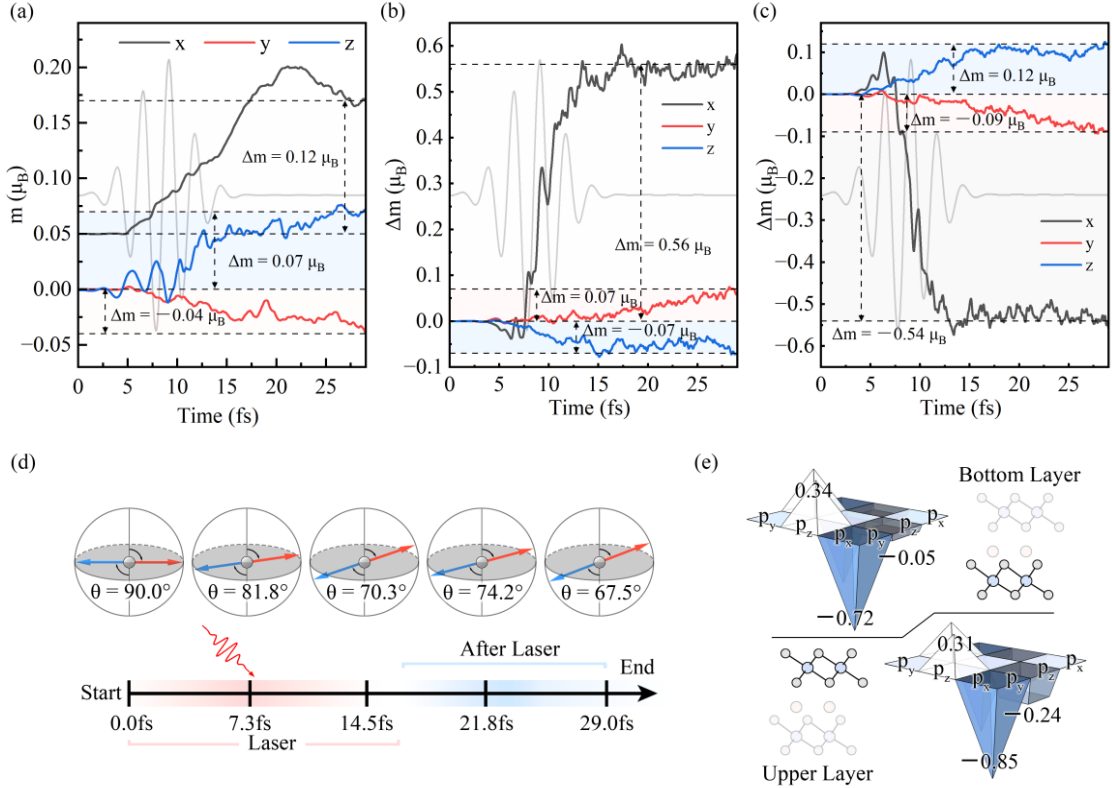

**Figure 3.** Time evolution of x-, y, and z-components of (a) total magnetic moment, (b) magnetic moment of upper $CrSe_2$ sublayer, and (c) magnetic moment of lower $CrSe_2$ sublayer under laser pulse. Gray lines in (a-c) represent the vector potential of the laser pulse. (d) Schematic diagram time evolution of spin slanting under laser pulse. Red and blue arrows in (d) represent the normalized magnetic moment of FE-1 and FE-2, respectively. (e) SOC matrixes of $p$ orbitals of Se atoms in two sublayers in the ground state, the unit of numbers is meV.

Along with the demagnetization process, the spin orientation of SL $CuCr_2Se_4$ is also changed. **Figure 3(a)** illustrates the evolution of x-, y-, and z-components of total magnetic moment SL $CuCr_2Se_4$ under laser pulse. Before imposing laser pulse, the y- and z-components of the magnetic moment is 0 $\mu_B$, corresponding to the spin orientation along x-direction. When applying ultrafast laser pulse, significant time-dependent changes of spin orientation are observed for the total magnetic moment. Specifically, the significant increase occurs for x- and z-components, which begins at ~ 4 fs and then stabilizes at ~ 26 fs. This results in the change of x/z-components of $\Delta m =$ ~ 0.12/0.07 $\mu_B$. While for y-component, it starts to magnetize from ~ 5 fs. In detail, y-component gains a magnetic moment of ~ - 0.04 $\mu_B$. Obviously, the laser-induced different responses of x-, y, and z-components imply the spin slanting in SL CuCr$_2$Se$_4$. To visually observe the spin slanting, the normalized magnetic moments are shown in **Figure 3(d)**. After applying ultrafast laser pulse, the spin is deflected by 22.5°. It should be noted that the small changes in the direction and magnitude of magnetic moments can be well measured and detected in experiment[49].

**Figures 3(b)** and **3(c)** present the time evolution of x-, y- and z-components of magnetic moments of upper and lower CrSe$_2$ sublayers under laser pulse. Clearly, the changes in x-, y- and z-components of the two CrSe$_2$ sublayers are different. For z-component, it is magnetized by 0.12 $\mu_B$ in lower CrSe$_2$ sublayer and magnetized by - 0.07 $\mu_B$ in upper CrSe$_2$ sublayer. While for the y-component, it is magnetized by - 0.09 $\mu_B$ in lower CrSe$_2$ sublayer and magnetized by 0.07 $\mu_B$ in upper CrSe$_2$ sublayer, which is almost canceled with each other. Therefore, as compared to light-induced magnetization in y-component, magnetization in z-component contributes more to the spin slanting in total magnetic moment.

Such light-induced spin slanting is expected from the proposed mechanism. Concerning SOC-induced interaction in SL CuCr$_2$Se$_4$, as shown in **Figure 3(e)** and **S2**, since Se atom is much heavier than Cr atom, similar to many previous studied systems[3,50,51], the main contribution of SOC-induced interaction does not originate from the magnetic atoms, but from its adjacent atoms. From **Figure 2(b)**, we can see that around the Fermi energy, on the Γ-K line, the $p_z$ orbital is occupied, and some of $p_x$ and $p_y$ orbitals are empty. This results in the SOC-induced interaction $\langle p_{x/y}|\hat{H}_{SOC}^0|p_z\rangle$ dominates the spin orientation, which favors in-plane, agreeing well with our calculations. For the lowest conduction states near the M-Γ path, they are mainly contributed by $p_x$ and $p_y$ orbitals of Se atoms and $d$ orbitals of Cr atoms (see **Figure S1**). After laser pulse pumping, as shown in **Figure 2(d)**, electrons are excited to these states, i.e., some of $p_x$ and $p_y$ orbitals in the conduction band are partially occupied. This introduces an extra SOC-induced interaction $\langle p_{x/y}|\hat{H}_{SOC}^0|p_{y/x}\rangle$, which prefers out-of-plane spin orientation. The competition between $\langle p_{x/y}|\hat{H}_{SOC}^0|p_z\rangle$ and $\langle p_{x/y}|\hat{H}_{SOC}^0|p_{y/x}\rangle$ causes the spin to deviate from its original direction in SL CuCr$_2$Se$_4$.

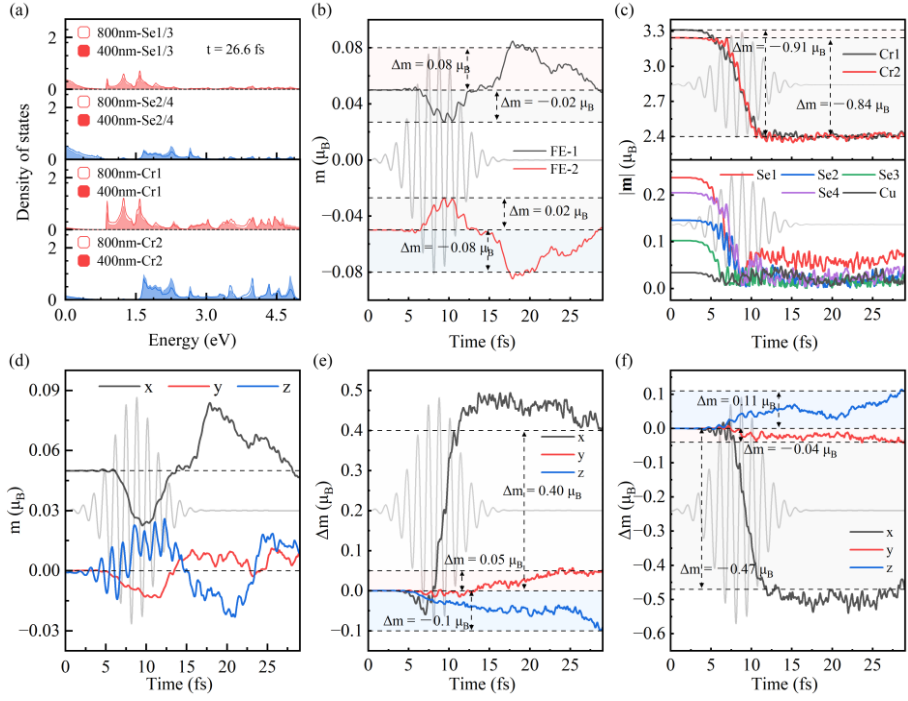

**Figure 4.** Results of ultrafast laser pulse with wavelength of 400 nm. (a) Orbital-resolved density of occupied states after laser excitation for SL CuCr$_2$Se$_4$ under FE-1. The lines indicate the density of occupied states under 800 nm laser pulse and the fill color region indicates the density of occupied states under 400 nm laser pulse. (b) Time-dependent dynamics of total magnetic moments of two ferroelectricity states. (c) Time evolution of local magnetic moment on each atom under laser excitation for SL CuCr$_2$Se$_4$ under FE-1. Time evolution of x-, y- and z-components of (d) total magnetic moment, (e) magnetic moment of upper CrSe$_2$ sublayer, and (f) magnetic moment of lower CrSe$_2$ sublayer under laser pulse. Gray line in (b-f) represents the absolute value of the vector potential of laser pulse.

To understand the effect of frequency of ultrafast laser pulse on magnetization dynamics in SL CuCr$_2$Se$_4$, we further apply a laser pulse with the same parameters as before but a wavelength of 400 nm on the system. With the photon energy of 3.10 eV, electrons can be excited to higher energy states. This will result in a different distribution of occupied states compared to the case of 800 nm. As shown in **Figure 4(a)**, under 400 nm laser pulse, more states above 1.5 eV are occupied, which are mainly contributed by Cr2, Se2, and Se4 atoms. The density of occupied states with energy near 0.8 eV is decreased, which are primarily contributed by Cr1, Se1, and Se3 atoms. Such differences induce different demagnetization behaviors and SOC-induced interactions in SL CuCr$_2$Se$_4$. As shown in **Figure 4(b)**, the magnetic moment first decreases ~ 0.02 μ$_B$ at the peak of the laser pulse (around the 8 fs), and then increases to 0.08 μ$_B$ at 18 fs. After 18 fs, the magnetic moment decreases to the initial magnetic moment at the end of the simulation. **Figure 4(c)** shows time evolution of local magnetic moment on each atom under laser excitation for SL CuCr$_2$Se$_4$. It can be seen that the magnetic moment difference between Cr1 and Cr2 atoms decreases from 0.07 $\mu_B$ to almost zero [see upper panel of **Figure 4(c)**]. However, as shown in the lower panel of **Figure 4(c)**, the magnetic moment on Se1 atom first decreases to almost zero at the peak of the laser pulse, and then increases to 0.08 $\mu_B$ in a few femtoseconds.

As discussed above, the changes in distribution of occupied states would introduce extra SOC-induced interactions, which might change the spin orientation. As shown in **Figure 4(d)**, the x-component of total magnetic moment varies similar to the total magnetic moment. While for the y- and z-components, magnetization is observed, exhibiting significant changes. From the time evolutions of x-, y- and z-components of magnetic moments of upper and lower CrSe$_2$ sublayers shown in **Figures 4(e)** and **(f)**, we can that the difference in changes of y/z-components of two CrSe$_2$ sublayers are relatively small as compared with the case of 800 nm. This gives rise to a relatively small spin slanting of 18° under the laser pulse. As a result, the spin slanting can be modulated by engineering the frequency of light.

At we wish to stress that as shown in **Figure 1(c)** and **4(b)**, when transforming into FE-2 phase, similar results are obtained. However, the spin slanting occurs in the opposite direction, as shown in **Figure 3(b)**. This indicate that the light-induced spin slanting in SL CuCr$_2$Se$_4$ can be coupled with ferroelectricity, generating the FE control of light-induced spin slanting.

## 3. Conclusion

In summary, based on the framework of light-induced magnetization dynamics, we report a mechanism of manipulating spins along directions beyond in-plane and out-of-plane in 2D multiferroic materials. Our straightforward perturbation theory analysis unveils that the competition between SOC-induced interactions from in-plane and out-of-plane orbitals can give rise to spin slanting. And such competition can be introduced through ultrafast light illumination as optical excitation can alter the orbital occupation, which leads to light-induced spin slanting. Furthermore, our rt-TDDFT also validates this mechanism in SL CuCr$_2$Se$_4$.

**Supplement Information**

(Table S1) Correlation between SOC-induced interactions of d orbitals and preferred spin orientation. (Figure S1) Orbital-resolved spin-polarization band structure of the FE-1 SL CuCr$_2$Se$_4$. (Figure S2) The SOC matrix of the *d* orbitals of Cr atoms. (Figure S3) The computational result of laser pulse having the intensity of $9\times10^{12}$ w/cm$^2$. (Note S1) Discussion of $9\times10^{12}$ w/cm$^2$ result. (Note S2) Computational method.

**Acknowledgment**

This work is supported by the National Natural Science Foundation of China (Nos. 12274261 and 12074217), Taishan Young Scholar Program of Shandong Province, and Shandong Provincial Natural Science Foundation of China (No. ZR2024QA016).

**Reference**

(1) Huang, B.; Clark, G.; Navarro-Moratalla, E.; Klein, D. R.; Cheng, R.; Seyler, K. L.; Zhong, D.; Schmidgall, E.; McGuire, M. A.; Cobden, D. H.; Yao, W.; Xiao, D.; Jarillo-Herrero, P.; Xu, X. Layer-Dependent Ferromagnetism in a van Der Waals Crystal down to the Monolayer Limit. *Nature* **2017**, *546* (7657), 270–273. https://doi.org/10.1038/nature22391.

(2) Gong, C.; Li, L.; Li, Z.; Ji, H.; Stern, A.; Xia, Y.; Cao, T.; Bao, W.; Wang, C.; Wang, Y.; Qiu, Z. Q.; Cava, R. J.; Louie, S. G.; Xia, J.; Zhang, X. Discovery of Intrinsic Ferromagnetism in Two-Dimensional van Der Waals Crystals. *Nature* **2017**, *546* (7657), 265–269. https://doi.org/10.1038/nature22060.

(3) Kim, J.; Kim, K.-W.; Kim, B.; Kang, C.-J.; Shin, D.; Lee, S.-H.; Min, B.-C.; Park, N. Exploitable Magnetic Anisotropy of the Two-Dimensional Magnet $CrI_3$. *Nano Lett.* **2020**, *20* (2), 929–935. https://doi.org/10.1021/acs.nanolett.9b03815.

(4) Ye, C.; Wang, C.; Wu, Q.; Liu, S.; Zhou, J.; Wang, G.; Söll, A.; Sofer, Z.; Yue, M.; Liu, X.; Tian, M.; Xiong, Q.; Ji, W.; Renshaw Wang, X. Layer-Dependent Interlayer Antiferromagnetic Spin Reorientation in Air-Stable Semiconductor CrSBr. *ACS Nano* **2022**, *16* (8), 11876–11883. https://doi.org/10.1021/acsnano.2c01151.

(5) Peng, R.; Ma, Y.; Xu, X.; He, Z.; Huang, B.; Dai, Y. Intrinsic Anomalous Valley Hall Effect in Single-Layer $Nb_3I_8$. *Phys. Rev. B* **2020**, *102* (3), 035412. https://doi.org/10.1103/PhysRevB.102.035412.

(6) Liang, Y.; Zhao, P.; Zheng, F.; Frauenheim, T. Ferroelectric Antiferromagnetic Quantum Anomalous Hall Insulator in Two-Dimensional van Der Waals Materials. *Phys. Rev. B* **2024**, *110* (20), 205421. https://doi.org/10.1103/PhysRevB.110.205421.

(7) Zhou, J.; Wang, Q.; Sun, Q.; Chen, X. S.; Kawazoe, Y.; Jena, P. Ferromagnetism in Semihydrogenated Graphene Sheet. *Nano Lett.* **2009**, *9* (11), 4.

(8) Yu, R.; Zhang, W.; Zhang, H.-J.; Zhang, S.-C.; Dai, X.; Fang, Z. Quantized Anomalous Hall Effect in Magnetic Topological Insulators. *Science* **2010**, *329* (5987), 61–64. https://doi.org/10.1126/science.1187485.

(9) Canonico, L. M.; Rappoport, T. G.; Muniz, R. B. Spin and Charge Transport of Multiorbital Quantum

Spin Hall Insulators. *Phys. Rev. Lett.* **2019**, *122* (19), 196601. https://doi.org/10.1103/PhysRevLett.122.196601.

(10) Antebi, O.; Stern, A.; Berg, E. Stoner Ferromagnetism in a Momentum-Confined Interacting 2D Electron Gas. *Phys. Rev. Lett.* **2024**, *132* (8), 086501. https://doi.org/10.1103/PhysRevLett.132.086501.

(11) Wang, X.; Xiao, C.; Park, H.; Zhu, J.; Wang, C.; Taniguchi, T.; Watanabe, K.; Yan, J.; Xiao, D.; Gamelin, D. R.; Yao, W.; Xu, X. Light-Induced Ferromagnetism in Moiré Superlattices. *Nature* **2022**, *604* (7906), 468–473. https://doi.org/10.1038/s41586-022-04472-z.

(12) Gao, A.; Liu, Y.-F.; Hu, C.; Qiu, J.-X.; Tzschaschel, C.; Ghosh, B.; Ho, S.-C.; Bérubé, D.; Chen, R.; Sun, H.; Zhang, Z.; Zhang, X.-Y.; Wang, Y.-X.; Wang, N.; Huang, Z.; Felser, C.; Agarwal, A.; Ding, T.; Tien, H.-J.; Akey, A.; Gardener, J.; Singh, B.; Watanabe, K.; Taniguchi, T.; Burch, K. S.; Bell, D. C.; Zhou, B. B.; Gao, W.; Lu, H.-Z.; Bansil, A.; Lin, H.; Chang, T.-R.; Fu, L.; Ma, Q.; Ni, N.; Xu, S.-Y. Layer Hall Effect in a 2D Topological Axion Antiferromagnet. *Nature* **2021**, *595* (7868), 521–525. https://doi.org/10.1038/s41586-021-03679-w.

(13) Chen, H.; Niu, Q.; MacDonald, A. H. Anomalous Hall Effect Arising from Noncollinear Antiferromagnetism. *Phys. Rev. Lett.* **2014**, *112* (1), 017205. https://doi.org/10.1103/PhysRevLett.112.017205.

(14) Wang, X.; Chen, A.; Wu, X.; Zhang, J.; Dong, J.; Zhang, L. Synthesis and Modulation of Low-Dimensional Transition Metal Chalcogenide Materials via Atomic Substitution. *Nano-Micro Lett.* **2024**, *16* (1), 163. https://doi.org/10.1007/s40820-024-01378-5.

(15) Zhang, C.; Zhang, L.; Tang, C.; Sanvito, S.; Zhou, B.; Jiang, Z.; Du, A. First-Principles Study of a Mn-Doped Monolayer: Coexistence of Ferromagnetism and Ferroelectricity with Robust Half-Metallicity and Enhanced Polarization. *Phys. Rev. B* **2020**, *102* (13), 134416. https://doi.org/10.1103/PhysRevB.102.134416.

(16) Zhou, Z.; Wang, H.; Li, X. Multiple Valley Modulations in Noncollinear Antiferromagnets. *Nano Lett.* **2024**, *24* (37), 11497–11503. https://doi.org/10.1021/acs.nanolett.4c02849.

(17) Kirilyuk, A.; Kimel, A. V.; Rasing, T. Ultrafast Optical Manipulation of Magnetic Order. *Rev. Mod. Phys.* **2010**, *82* (3), 2731–2784. https://doi.org/10.1103/RevModPhys.82.2731.

(18) Bao, D.-L.; O'Hara, A.; Du, S.; Pantelides, S. T. Tunable, Ferroelectricity-Inducing, Spin-Spiral Magnetic Ordering in Monolayer FeOCl. *Nano Lett.* **2022**, *22* (9), 3598–3603. https://doi.org/10.1021/acs.nanolett.1c05043.

(19) Zhang, D.; Li, A.; Chen, X.; Zhou, W.; Ouyang, F. Tuning Valley Splitting and Magnetic Anisotropy of

Multiferroic $CuMP_2X_6$ ( M = Cr , V ; X = S , Se ) Monolayer. *Phys. Rev. B* **2022**, *105* (8), 085408. https://doi.org/10.1103/PhysRevB.105.085408.

(20) Krempaský, J.; Šmejkal, L.; D'Souza, S. W.; Hajlaoui, M.; Springholz, G.; Uhlířová, K.; Alarab, F.; Constantinou, P. C.; Strocov, V.; Usanov, D.; Pudelko, W. R.; González-Hernández, R.; Birk Hellenes, A.; Jansa, Z.; Reichlová, H.; Šobáň, Z.; Gonzalez Betancourt, R. D.; Wadley, P.; Sinova, J.; Kriegner, D.; Minár, J.; Dil, J. H.; Jungwirth, T. Altermagnetic Lifting of Kramers Spin Degeneracy. *Nature* **2024**, *626* (7999), 517–522. https://doi.org/10.1038/s41586-023-06907-7.

(21) Železný, J.; Zhang, Y.; Felser, C.; Yan, B. Spin-Polarized Current in Noncollinear Antiferromagnets. *Phys. Rev. Lett.* **2017**, *119* (18), 187204. https://doi.org/10.1103/PhysRevLett.119.187204.

(22) Song, Q.; Occhialini, C. A.; Ergeçen, E.; Ilyas, B.; Amoroso, D.; Barone, P.; Kapeghian, J.; Watanabe, K.; Taniguchi, T.; Botana, A. S.; Picozzi, S.; Gedik, N.; Comin, R. Evidence for a Single-Layer van Der Waals Multiferroic. *Nature* **2022**, *602* (7898), 601–605. https://doi.org/10.1038/s41586-021-04337-x.

(23) Li, J.; Li, Y.; Du, S.; Wang, Z.; Gu, B.-L.; Zhang, S.-C.; He, K.; Duan, W.; Xu, Y. Intrinsic Magnetic Topological Insulators in van Der Waals Layered $MnBi_2Te_4$-Family Materials. *Sci. Adv.* **2019**, *5* (6), eaaw5685. https://doi.org/10.1126/sciadv.aaw5685.

(24) Bai, J.; Yang, T.; Guo, Z.; Liu, Y.; Jiao, Y.; Meng, W.; Cheng, Z. Controllable Topological Phase Transition via Ferroelectric–Paraelectric Switching in a Ferromagnetic Single-Layer $M_IM_{II}Ge_2X_6$ Family. *Mater. Horiz.* **2025**. https://doi.org/10.1039/D4MH01599A.

(25) He, Z.; Peng, R.; Feng, X.; Xu, X.; Dai, Y.; Huang, B.; Ma, Y. Two-Dimensional Valleytronic Semiconductor with Spontaneous Spin and Valley Polarization in Single-Layer $Cr_2Se_3$. *Phys. Rev. B* **2021**, *104* (7), 075105. https://doi.org/10.1103/PhysRevB.104.075105.

(26) Xue, Q.; Sun, Y.; Zhou, J. Nonlinear Optics-Driven Spin Reorientation in Ferromagnetic Materials. *ACS Nano* **2024**, *18* (35), 24317–24326. https://doi.org/10.1021/acsnano.4c06453.

(27) Duan, X.; Wang, H.; Chen, X.; Qi, J. Multiple Polarization Phases and Strong Magnetoelectric Coupling in the Layered Transition Metal Phosphorus Chalcogenides $TMP_2X_6$ ( T = Cu , Ag ; M = Cr , V ; X = S , Se ) by Controlling the Interlayer Interaction and Dimension. *Phys. Rev. B* **2022**, *106* (11), 115403. https://doi.org/10.1103/PhysRevB.106.115403.

(28) Zhang, J.; Tancogne-Dejean, N.; Xian, L.; Boström, E. V.; Claassen, M.; Kennes, D. M.; Rubio, A. Ultrafast Spin Dynamics and Photoinduced Insulator-to-Metal Transition in α-$RuCl_3$. *Nano Lett.* **2023**, *23* (18), 8712–8718. https://doi.org/10.1021/acs.nanolett.3c02668.

(29) Liu, X.; Legut, D.; Zhang, Q. Light-Induced Ultrafast Enhancement of Magnetic Orders in Monolayer

CrX$_3$. *J. Phys. Chem. C* **2023**, acs.jpcc.3c03187. https://doi.org/10.1021/acs.jpcc.3c03187.

(30) Conte, F.; Ninno, D.; Cantele, G. Layer-Dependent Electronic and Magnetic Properties of Nb$_3$I$_8$. *Phys. Rev. Res.* **2020**, *2* (3), 033001. https://doi.org/10.1103/PhysRevResearch.2.033001.

(31) Burch, K. S.; Mandrus, D.; Park, J.-G. Magnetism in Two-Dimensional van Der Waals Materials. *Nature* **2018**, *563* (7729), 47–52. https://doi.org/10.1038/s41586-018-0631-z.

(32) Afanasiev, D.; Hortensius, J. R.; Matthiesen, M.; Mañas-Valero, S.; Šiškins, M.; Lee, M.; Lesne, E.; Zant, H. S. J. van der; Steeneken, P. G.; Ivanov, B. A.; Coronado, E.; Caviglia, A. D. Controlling the Anisotropy of a van Der Waals Antiferromagnet with Light. *Sci. Adv.* **2021**, *7* (23), eabf3096. https://doi.org/10.1126/sciadv.abf3096.

(33) Whangbo, M.-H.; Gordon, E. E.; Xiang, H.; Koo, H.-J.; Lee, C. Prediction of Spin Orientations in Terms of HOMO–LUMO Interactions Using Spin–Orbit Coupling as Perturbation. *Acc. Chem. Res.* **2015**, *48* (12), 3080–3087. https://doi.org/10.1021/acs.accounts.5b00408.

(34) Whangbo, M.-H.; Xiang, H.; Koo, H.-J.; Gordon, E. E.; Whitten, J. L. Electronic and Structural Factors Controlling the Spin Orientations of Magnetic Ions. *Inorg. Chem.* **2019**, *58* (18), 11854–11874. https://doi.org/10.1021/acs.inorgchem.9b00687.

(35) Yang, B. S.; Zhang, J.; Jiang, L. N.; Chen, W. Z.; Tang, P. Strain Induced Enhancement of Perpendicular Magnetic Anisotropy in Co/Graphene and Co/BN Heterostructures. *Phys. Rev. B* **2017**.

(36) Bhattacharya, S.; Basu, R.; Bhatt, R.; Pitale, S.; Singh, A.; Aswal, D. K.; Gupta, S. K.; Navaneethan, M.; Hayakawa, Y. CuCrSe2: A High Performance Phonon Glass and Electron Crystal Thermoelectric Material. *J. Mater. Chem. A* **2013**, *1* (37), 11289–11294. https://doi.org/10.1039/C3TA11903C.

(37) Peng, J.; Su, Y.; Lv, H.; Wu, J.; Liu, Y.; Wang, M.; Zhao, J.; Guo, Y.; Wu, X.; Wu, C.; Xie, Y. Even–Odd-Layer-Dependent Ferromagnetism in 2D Non-van-Der-Waals CrCuSe$_2$. *Adv. Mater.* **2023**, *35* (16), 2209365. https://doi.org/10.1002/adma.202209365.

(38) Yan, Y.; Guo, L.; Zhang, Z.; Lu, X.; Peng, K.; Yao, W.; Dai, J.; Wang, G.; Zhou, X. Sintering Temperature Dependence of Thermoelectric Performance in CuCrSe$_2$ Prepared via Mechanical Alloying. *Scr. Mater.* **2017**, *127*, 127–131. https://doi.org/10.1016/j.scriptamat.2016.09.016.

(39) Cheng, Y.; Yang, J.; Jiang, Q.; Fu, L.; Xiao, Y.; Luo, Y.; Zhang, D.; Zhang, M. CuCrSe$_2$ Ternary Chromium Chalcogenide: Facile Fabrication, Doping and Thermoelectric Properties. *J. Am. Ceram. Soc.* **2015**, *98* (12), 3975–3980. https://doi.org/10.1111/jace.13860.

(40) Sun, Z.; Su, Y.; Zhi, A.; Gao, Z.; Han, X.; Wu, K.; Bao, L.; Huang, Y.; Shi, Y.; Bai, X.; Cheng, P.; Chen, L.; Wu, K.; Tian, X.; Wu, C.; Feng, B. Evidence for Multiferroicity in Single-Layer CuCrSe$_2$. *Nat.*

*Commun.* **2024**, *15* (1), 4252. https://doi.org/10.1038/s41467-024-48636-z.

(41) Zhang, Y.; Hu, Y.; Deng, Z.; Lou, Z.; Hou, Y.; Teng, F. 2D Growth of High-Quality Magnetic CuCr$_2$Se$_4$ Crystals with a High Curie Temperature. *Mater. Lett.* **2022**, *310*, 131478. https://doi.org/10.1016/j.matlet.2021.131478.

(42) Dou, K.; He, Z.; Zhao, J.; Du, W.; Dai, Y.; Huang, B.; Ma, Y. Engineering Topological Spin Hall Effect in 2D Multiferroic Material. *Adv. Sci.* **2024**, 2407982. https://doi.org/10.1002/advs.202407982.

(43) Yang, L.; Li, L.; Yu, Z.-M.; Wu, M.; Yao, Y. Two-Dimensional Topological Ferroelectric Metal with Giant Shift Current. *Phys. Rev. Lett.* **2024**, *133* (18), 186801. https://doi.org/10.1103/PhysRevLett.133.186801.

(44) Fei, Z.; Zhao, W.; Palomaki, T. A.; Sun, B.; Miller, M. K.; Zhao, Z.; Yan, J.; Xu, X.; Cobden, D. H. Ferroelectric Switching of a Two-Dimensional Metal. *Nature* **2018**, *560* (7718), 336–339. https://doi.org/10.1038/s41586-018-0336-3.

(45) Lu, J.; Chen, G.; Luo, W.; Íñiguez, J.; Bellaiche, L.; Xiang, H. Ferroelectricity with Asymmetric Hysteresis in Metallic LiOsO$_3$ Ultrathin Films. *Phys. Rev. Lett.* **2019**, *122* (22), 227601. https://doi.org/10.1103/PhysRevLett.122.227601.

(46) Xu, W.; Shao, Y.-P.; Wang, J.-L.; Zheng, J.-D.; Tong, W.-Y.; Duan, C.-G. Origin of Metallic Ferroelectricity in Group-V Monolayer Black Phosphorus. *Phys. Rev. B* **2024**, *109* (3), 035421. https://doi.org/10.1103/PhysRevB.109.035421.

(47) Du, R.; Wang, Y.; Cheng, M.; Wang, P.; Li, H.; Feng, W.; Song, L.; Shi, J.; He, J. Two-Dimensional Multiferroic Material of Metallic p-Doped SnSe. *Nat. Commun.* **2022**, *13* (1), 6130. https://doi.org/10.1038/s41467-022-33917-2.

(48) He, J.; Li, S.; Frauenheim, T.; Zhou, Z. Ultrafast Laser Pulse Induced Transient Ferrimagnetic State and Spin Relaxation Dynamics in Two-Dimensional Antiferromagnets. *Nano Lett.* **2023**, *23* (17), 8348–8354. https://doi.org/10.1021/acs.nanolett.3c02727.

(49) Deng, S.; Gomonay, O.; Chen, J.; Fischer, G.; He, L.; Wang, C.; Huang, Q.; Shen, F.; Tan, Z.; Zhou, R.; Hu, Z.; Šmejkal, L.; Sinova, J.; Wernsdorfer, W.; Sürgers, C. Phase Transitions Associated with Magnetic-Field Induced Topological Orbital Momenta in a Non-Collinear Antiferromagnet. *Nat. Commun.* **2024**, *15* (1), 822. https://doi.org/10.1038/s41467-024-45129-x.

(50) Tran, H. B.; Momida, H.; Matsushita, Y.; Shirai, K.; Oguchi, T. Insight into Anisotropic Magnetocaloric Effect of CrI$_3$. *Acta Mater.* **2022**, *231*, 117851. https://doi.org/10.1016/j.actamat.2022.117851.

(51) Zhu, H.; Gao, Y.; Hou, Y.; Gui, Z.; Huang, L. Tunable Magnetic Anisotropy in Two-Dimensional

CrX$_3$/AlN ( X= I, Br, Cl ) Heterostructures. *Phys. Rev. B* **2022**, *106* (13), 134412. https://doi.org/10.1103/PhysRevB.106.134412.